\documentclass[aps,prd,a4paper,amsmath,amssymb,nofootinbib,showpacs,showkeys,notitlepage,superscriptaddress]{revtex4-1}

\usepackage{bm}
\usepackage[dvips]{color,graphicx}

\begin{document}

\title{Phase diagram of hot quark matter under magnetic field}


\author{Marco Ruggieri}\email{ruggieri@yukawa.kyoto-u.ac.jp}
\affiliation{Yukawa Institute for Theoretical Physics,
 Kyoto University, Kyoto 606-8502, Japan}

\begin{abstract}
In this talk, I review the computation of the phase diagram of hot
quark matter in strong magnetic field, at zero baryon density,
within an effective model of Quantum Chromodynamics.
\end{abstract}

\maketitle


\section{Introduction}
The modification of the QCD vacuum, and of its thermal excitations
as well, under the influence of external fields, is an attractive
topic. Firstly, it is extremely interesting to understand how an
external field can modify the main characteristics of confinement
and spontaneous chiral symmetry breaking. Secondly, strong
magnetic fields, with order of magnitude between $eB\approx
m_\pi^2$ and $eB\approx 15 m_\pi^2$, might be produced in the very
first moments of the non-central heavy ion
collisions~\cite{Kharzeev:2007jp,Skokov:2009qp}. In this case, it
has been argued that the non-trivial topological structure of
thermal QCD gives rise to Chiral Magnetic Effect (CME)
\cite{Kharzeev:2007jp,Buividovich:2009wi,Fukushima:2008xe}.

An useful approach to the physics of strong interactions in
external magnetic fields is the use of some model. Among them, the
Nambu-Jona Lasinio (NJL) model~\cite{Nambu:1961tp} is quite
popular, see Refs.~\cite{revNJL} for reviews. In this model, the
QCD gluon-mediated interactions are replaced by effective
interactions among quarks, which are built in order to respect the
global symmetries of QCD. On the other hand, the NJL model lacks
confinement of color. It is well known that color confinement can
be described in terms of the center symmetry of the color gauge
group and of the Polyakov loop~\cite{Polyakovetal}, which is an
order parameter for the center symmetry. Motivated by this
property, the Polyakov loop extended Nambu-Jona Lasinio model
(P-NJL model) has been
introduced~\cite{Meisinger:1995ih,Fukushima:2003fw}, in which the
concept of statistical confinement replaces that of the true
confinement of QCD, and an effective interaction among the chiral
condensate and the Polyakov loop is achieved by a covariant
coupling of quarks with a background temporal gluon field. In the
literature, there are several studies about various aspects of the
P-NJL
model~\cite{Ratti:2005jh,Roessner:2006xn,Megias:2006bn,Sasaki:2006ww,Ghosh:2007wy,
Fukushima:2008wg,Fukushima:2008wg,Abuki:2008nm,Kahara:2010wh,Sakai:2008py,Sakai:2009dv,
Kashiwa:2009ki,Abuki:2008tx,Sasaki:2010jz,Hell:2008cc,Kashiwa:2007hw,Gatto:2010qs}.
Lattice studies on the response of the QCD ground state to
external magnetic and chromomagnetic fields can be found
in~\cite{D'Elia:2010nq,Buividovich:2009my,Buividovich:2008wf,Cea:2002wx,Cea:2007yv}.
Previous studies of QCD in magnetic fields, and of QCD-like
theories as well, can be found in
Refs.~\cite{Klevansky:1989vi,Gusynin:1995nb,Klimenko:1990rh,Agasian:2008tb,
Fukushima:2010fe,Mizher:2010zb,Campanelli:2009sc}.

Beside the Polyakov loop, it has been
suggested~\cite{Bilgici:2008qy} that another observable which is
an order parameter for the center symmetry, hence for confinement,
is the dressed Polyakov loop. The dressed Polyakov loop has been
computed in Refs.~\cite{Fischer:2009wc} within the scheme of
truncated Schwinger-Dyson equations; within the Nambu-Jona Lasinio
model, $\Sigma_1$ has been computed at finite temperature and
chemical potential in~\cite{Mukherjee:2010cp}. Finally, the
dressed Polyakov loop has been computed within the PNJL model
in~\cite{Kashiwa:2009ki} at finite temperature, and
in~\cite{Gatto:2010qs} at finite temperature with strong magnetic
field.

In this talk, I present results obtained in
Ref.~\cite{Gatto:2010qs} about the phase structure and the dressed
Polyakov loop of hot two massive flavor quark matter at zero
chemical potential, in an external magnetic field. To compute the
effective potential, I rely on the PNJL model of strongly
interacting quarks.

\section{Theoretical framework}
I consider here two flavor quark matter whose Lagrangian density
is specified as~\cite{Gatto:2010qs}
\begin{eqnarray}
{\cal L} &=& \bar q\left(i\gamma^\mu D_\mu - m_0\right)q
        + g_\sigma\left[(\bar q q)^2 + (\bar q i \gamma_5 \bm\tau q)^2\right] \nonumber \\
        &&+g_8\left[(\bar q q)^2 + (\bar q i \gamma_5 \bm\tau
        q)^2\right]^2~.
\label{eq:lagr}
\end{eqnarray}
The covariant derivative embeds the quark coupling to the external
magnetic field and to the background gluon field as well, see
below. In Eq.~\eqref{eq:lagr}, $q$ represents a quark field in the
fundamental representation of color and flavor (indices are
suppressed for notational simplicity); $\bm\tau$ is a vector of
Pauli matrices in flavor space; $m_0$ is the bare quark mass,
which is fixed to reproduce the pion mass in the vacuum, $m_\pi =
139$ MeV. The model at hand is called Polyakov loop extended
Nambu-Jona Lasinio model (PNJL in the following), since a coupling
of the chiral condensate and the Polyakov loop is introduced via
the covariant derivative in Eq.~\eqref{eq:lagr}, see below.

In this study, I limit myself to the one-loop approximation for
the partition function. In order to couple the Polyakov loop to
the quark fields, it is customary, in the PNJL model, to introduce
a background temporal, static and homogeneous Euclidean gluon
field, $A_4$, in terms of which the Polyakov loop is given by $P =
\text{Tr}[\exp(i\beta A_4)]/3$. $A_4$ is coupled to the quarks via
the covariant derivative, see Eq.~(1); as a consequence, a
coupling among the quark fields and the Polyakov loop arises
naturally when the integration over fermion fields in the
partition function is performed.

I work in the Landau gauge, and take the magnetic field
homogeneous, static and aligned with the positive $z-$axe.
Moreover, I take twisted fermion boundary conditions along the
compact temporal direction,
\begin{equation}
q(\bm x,\beta) = e^{-i\varphi}q(\bm x,0)~,~~~\varphi\in[0,2\pi]~,
\label{eq:phi}
\end{equation}
while for spatial directions the usual periodic boundary condition
is taken. The one-loop thermodynamic potential in the general case
of twisted boundary conditions is given by~\cite{Gatto:2010qs}
\begin{eqnarray}
\Omega &=& {\cal U}(P,\bar P, T) + \frac{\sigma^2}{g_\sigma} +
\frac{3\sigma^4 g_8}{g_\sigma^4}
        - \sum_{f=u,d}\frac{|q_f e B|}{2\pi}
           \sum_{k}\alpha_{k}\int_{-\infty}^{+\infty}\frac{dp_z}{2\pi}g_\Lambda(p_z,k)\omega_{k}(p_z)
  \nonumber \\
      &-&T\sum_{f=u,d}\frac{|q_f e B|}{2\pi}\sum_{k}\alpha_{k}
        \int_{-\infty}^{+\infty}\frac{dp_z}{2\pi}
        \log\left(1+3P e^{-\beta\cal{E}_-} + 3\bar{P}e^{-2\beta\cal{E}_-}+e^{-3\beta\cal{E}_-} \right)
      \nonumber \\
     &-&T\sum_{f=u,d}\frac{|q_f e B|}{2\pi}\sum_{k}\alpha_{k}
        \int_{-\infty}^{+\infty}\frac{dp_z}{2\pi}
        \log\left(1+3\bar P e^{-\beta\cal{E}_+} + 3Pe^{-2\beta\cal{E}_+}+e^{-3\beta\cal{E}_+} \right)~.
\label{eq:Om1}
\end{eqnarray}

In the previous equation, $\sigma=g_\sigma\langle\bar q q\rangle =
2g_\sigma\langle\bar u u\rangle$; $k$ is a non-negative integer
which labels the Landau level; $\alpha_k = \delta_{k0} +
2(1-\delta_{k0})$ counts the degeneracy of the $k-$th Landau
level. Moreover, $\omega_k(p_z)^2=p_z^2 + 2|q_f eB| k + M^2$, with
$M = m_0 - 2\sigma -4\sigma^3 g_8/g_\sigma^3$~. The arguments of
the thermal exponentials are defined as ${\cal E}_\pm =
\omega_k(p_z) \pm i(\varphi-\pi)/\beta$, with $\varphi$ defined in
Eq.~\eqref{eq:phi}. The vacuum part of the thermodynamic
potential, $\Omega(T=0)$, is ultraviolet divergent.  In this
study, I use a smooth regularization procedure by introducing a
form factor $g_\Lambda(p)$ in the diverging zero-point energy. The
potential term $\mathcal{U}[P,\bar P,T]$ in Eq.~\eqref{eq:Om1} is
built by hand in order to reproduce the pure gluonic lattice
data~\cite{Ratti:2005jh}.  Among several different potential
choices I adopt the logarithmic form of~\cite{Ratti:2005jh}.

Following~\cite{Bilgici:2008qy}, I introduce the dual quark
condensate,
\begin{equation}
\tilde\Sigma_n(m,V) =
\int_0^{2\pi}\frac{d\varphi}{2\pi}\frac{e^{-i\varphi n}}{V}
\langle\bar q q\rangle_G~, \label{eq:Sn0}
\end{equation}
where $n$ is an integer. The expectation value
$\langle\bm\cdot\rangle_G$ denotes the path integral over gauge
field configurations. The case $n=1$ is called the {\em dressed}
Polyakov loop. For my later convenience, I scale the definition of
the dressed Polyakov loop in Eq.~\eqref{eq:Sn0}, and introduce
\begin{eqnarray}
\Sigma_1 &=& -2\pi
g_\sigma\int_0^{2\pi}\frac{d\varphi}{2\pi}e^{-i\varphi}
\langle\bar q q\rangle_G~, \nonumber \\
 &=& -
\int_0^{2\pi}\!d\varphi~e^{-i\varphi}
\sigma(\varphi)~,\label{eq:SnPNJL}
\end{eqnarray}
where $\sigma(\varphi)$ corresponds to the expectation value of
the $\sigma$ field computed keeping twisted boundary conditions
for fermions.

\section{Results}

\begin{figure}[t]
\includegraphics[width=8cm]{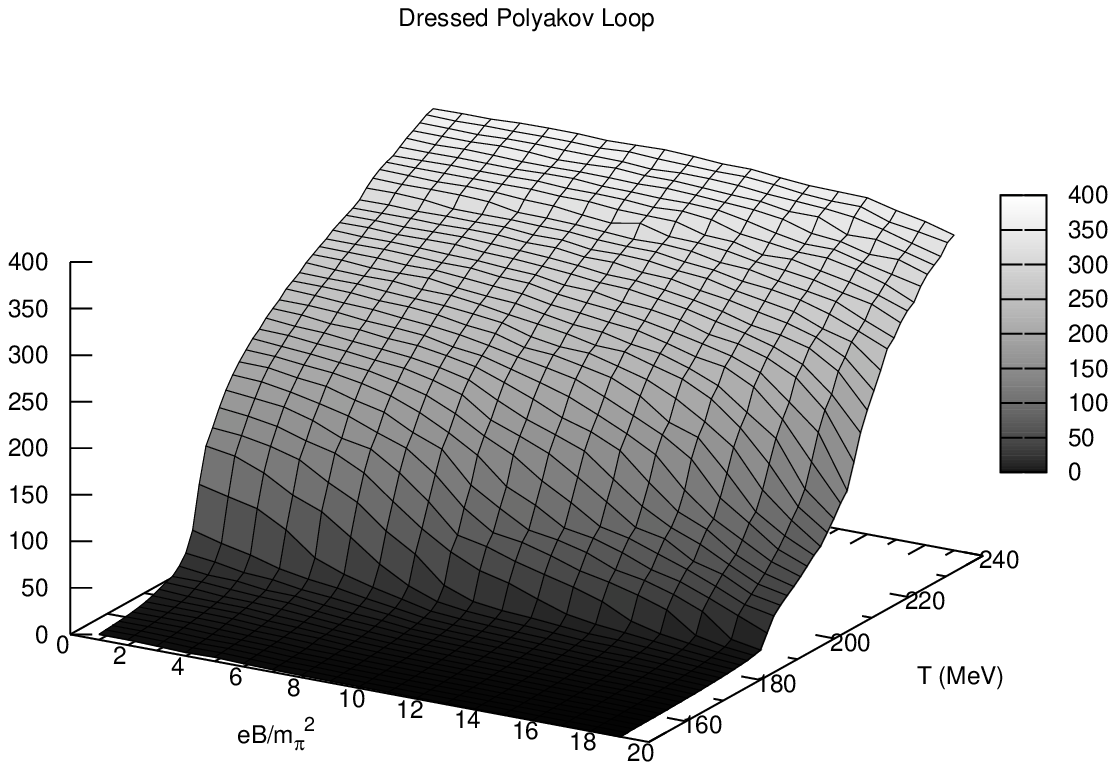}~~~~~\includegraphics[width=8cm]{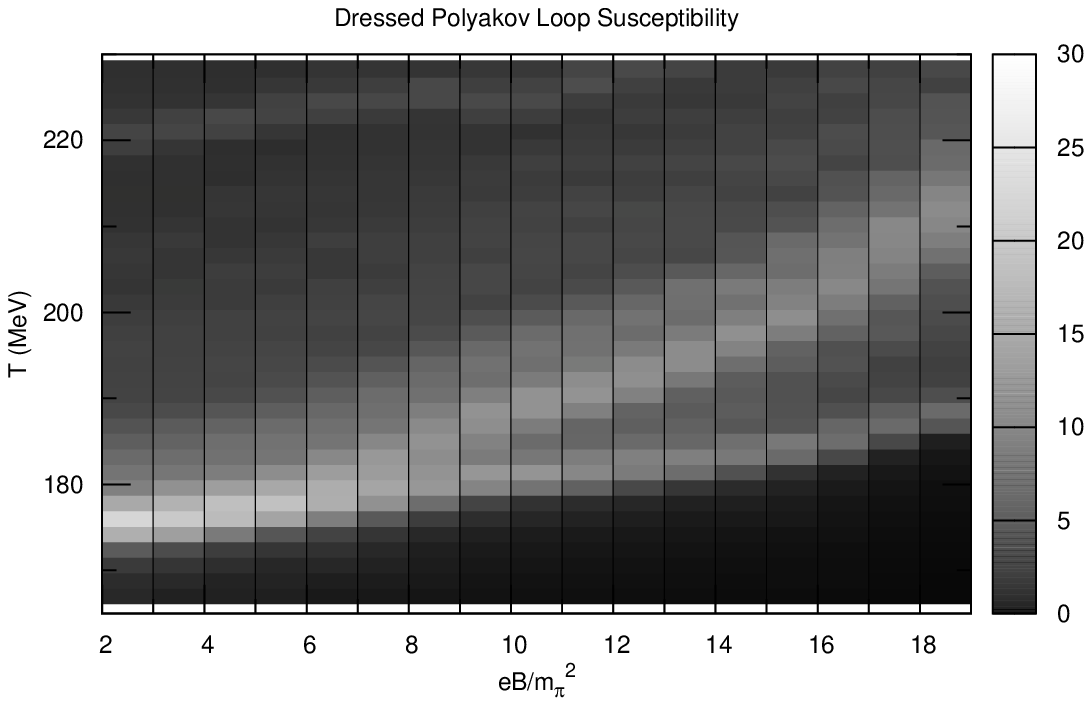}
\caption{\label{Fig:DPm} Dressed Polyakov loop (left panel) and
its effective susceptibility (right panel) as a function of
temperature and magnetic field strength.}
\end{figure}

In the left panel of Fig.~\ref{Fig:DPm}, I collect the results for
the dressed Polyakov loop as a function of temperature and
magnetic field strength. In the right panel of the same figure, I
plot the data of the effective susceptibility, $d\Sigma_1/dT$. The
bifurcation of the effective susceptibility at large value of the
magnetic field strength arises because the dressed Polyakov loop
is capable to feel both the deconfinement and the chiral
crossovers, which are split of almost $15\%$ if the magnetic field
strength is large enough.

\begin{figure}[t]
\includegraphics[width=8cm]{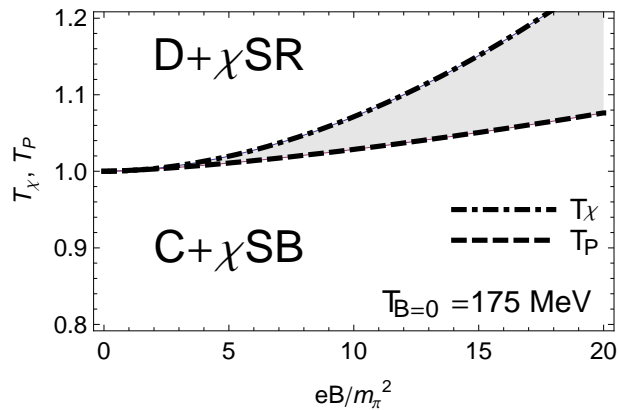}
\caption{\label{Fig:pd7} Phase diagram of the PNJL model in
magnetic field. Dashed line denotes the Polyakov loop crossover;
dot-dashed line corresponds to the chiral crossover. The shaded
area is the region, in the $eB-T$ plane, in which quark matter is
not statistically confined, but chiral symmetry is still broken by
the chiral condensate. Temperatures on the vertical axes are
measured in units of the pseudo-critical temperature at zero
field, which is $T_0 = 175$ MeV. }
\end{figure}

In Figure~\ref{Fig:pd7}, I collect the results on the
pseudo-critical temperatures for chiral and Polyakov loop
crossovers, in the form of phase diagrams in the $eB-T$ plane. The
dashed line denotes the Polyakov loop crossover, and the
dot-dashed line corresponds to the chiral crossover. The shaded
area is the region, in the $eB-T$ plane, in which quark matter is
not statistically confined, but chiral symmetry is still broken by
the chiral condensate. Temperature on the vertical axes are
measured in units of the pseudo-critical temperature at zero
field, which is $T_0 = 175$ MeV. We fit our data on the
pseudo-critical temperatures by the law
\begin{equation}
\frac{T_c^{A}}{T_0} = 1 + a\left(\frac{eB}{T_0}\right)^\alpha~,
\label{eq:fit}
\end{equation}
where $A=\sigma,P$, and $T_0 = 175$ MeV. The numerical values for
$T_\chi$ are $a=2.4\times 10^{-3}$, $\alpha = 1.85$; for $T_P$ the
numerical coefficients are $a=2.1\times 10^{-3}$, $\alpha = 1.41$.

It is instructive to compare this result with those obtained in a
different model. The shape of the phase diagram drawn in
Fig.~\ref{Fig:pd7} is similar to that drawn by the Polyakov
extended quark-meson model, see e.g. Fig.~13 of
Ref.~\cite{Mizher:2010zb}. In that reference, an interpretation of
the split in terms of the interplay among vacuum and thermal
contribution, is given. I totally agree with those arguments,
which are reproduced within the PNJL model as well, as the results
on critical temperatures show.

\section{Conclusions}
In this talk, I have reported about the computation of the dressed
Polyakov loop and of the phase diagram of hot quark matter under
the influence of a strong external magnetic field. The results on
the dressed Polyakov loop, $\Sigma_1$, in magnetic field show that
this quantity is capable to describe both Polyakov loop and chiral
crossovers. This is resumed in the double peak structure of its
effective susceptibility. We measure an increase of both
deconfinement and chiral crossovers; the tiny split of the two
critical temperatures is of the order of $10\%$ for the largest
value of the magnetic field strength considered here.

I acknowledge the organizers of QCD@Work10 for the opportunity to
give a talk. I acknowledge R.~Gatto for his fruitful collaboration
in Ref.~\cite{Gatto:2010qs}, on which this talk is mainly based.
Moreover, I acknowledge correspondence with M. Huang and
S.~Nicotri, and in particular with M. d'Elia. Furthermore, I
acknowledge stimulating discussions with L.~Campanelli and
K.~Fukushima. This work is supported by JSPS under the contract
number P09028. The numerical calculations were carried out on
Altix3700 BX2 at YITP in Kyoto University.



\bibliographystyle{aipproc}   

\end{document}